\documentclass[journal=jacsat,manuscript=article]{achemso}

\usepackage[version=3]{mhchem} 
\usepackage{mathtools}

\DeclarePairedDelimiter\ket{\lvert}{\rangle}
\DeclarePairedDelimiterX\braket[2]{\langle}{\rangle}{#1\,\delimsize\vert\,\mathopen{}#2}
\usepackage{tikz}
\usetikzlibrary{quantikz2}

\author{Michal Krompiec}
\email{michal.krompiec@fujitsu.com}
\author{Josh J. M. Kirsopp}
\author{Antonio M{\'a}rquez Romero}
\author{Vicente P. Soloviev}

\affiliation[Fujitsu]
{Fujitsu Research of Europe Ltd., Slough SL1 2BE, UK}

\title[UpCCD state prep]
  {A simple method for seniority-zero quantum state preparation}

\begin{document}

\begin{tocentry}
\includegraphics[width=\textwidth]{"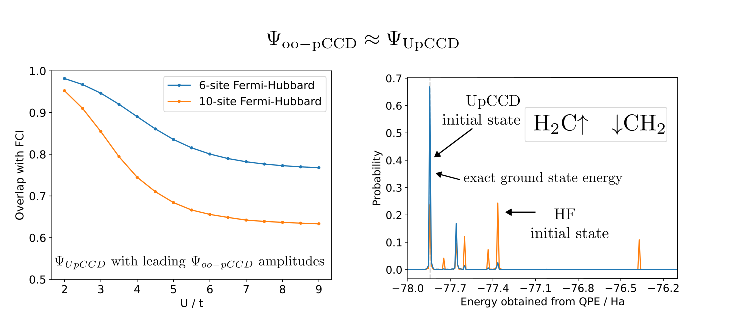"}
\end{tocentry}

\begin{abstract}
Quantum Phase Estimation (QPE), the quantum algorithm for estimating eigenvalues of a given Hermitian matrix and preparing its eigenvectors, is considered the most promising approach to finding the ground states and their energies of electronic systems using a quantum computer. It requires, however, to be warm--started from an initial state with sufficiently high overlap with the ground state. For strongly-correlated states, where QPE is expected to have advantage over classical methods, preparation of such initial states requires deep quantum circuits and/or expensive hybrid quantum-classical optimization. It is well-known that orbital-optimized paired Coupled Cluster Doubles (oo-pCCD) method can describe the static correlation features of many strongly correlated singlet states. We show that pCCD and its unitary counterpart, UpCCD, become equivalent in the limit of small amplitudes or if the number of large amplitudes is below 5. We demonstrate that substituting leading oo-pCCD amplitudes into the UpCCD Ansatz allows to prepare high-fidelity singlet states for models of multiple-bond dissociation in ethene, ethyne and dinitrogen, as well as for 1D Hubbard models at half-filling, with very shallow circuits. We envisage our method to be of general use for approximate preparation of singlet states for Quantum Phase Estimation and related algorithms.
\end{abstract}

\section{Introduction}
Quantum Phase Estimation (QPE) \cite{kitaev1995quantum,abrams1999quantum} is widely regarded as one of the most promising proposals for finding eigenvalues and eigenvectors of a given Hamiltonian using a quantum computer, particularly in cases where classical computing methods become computationally prohibitive. However, successful execution of QPE crucially depends on the value of the overlap $S$ of the \textit{initial state} it acts upon with the target state being large enough, as the cost of performing canonical QPE is proportional to $S^{-1}$ \cite{lee2023evaluating}, while more resource-efficient protocols have stricter requirements:  Quantum Complex Exponential Least Squares requires $S^2\ge0.71$ \cite{ding2023even}, Quantum Multiple Eigenvalue Gaussian filtered Search (when applied to the ground state) requires $S^2\ge0.5$ \cite{ding2024quantum}. While a trivial initial state, e.g. the Hartree-Fock wavefunction, may often have a high overlap with the ground state of closed-shell molecules at their equilibrium geometries (see e.g. the G1 test set \cite{tubman2018postponing}), this is clearly not the case for strongly-correlated or `multi-reference' systems. For these, several authors \cite{feniou2024sparse, tubman2018postponing, fomichev2024initial} suggest performing an approximate Full Configuration Interaction (Full CI) calculation (such as Adaptive Sampling CI, ASCI, \cite{tubman2016deterministic} Stochastic Heat-bath CI, SHCI \cite{sharma2017semistochastic} or CI Perturbatively Selected Iteratively, CIPSI \cite{huron1973iterative}) on a classical computer and then encoding the CI state in a quantum circuit, using e.g. QROM \cite{babbush2018encoding} or a specialized method for sparse states, such as CVO-QRAM \cite{deveras2022double}. It was also proposed \cite{fomichev2024initial, berry2025rapid} to start with a Density Matrix Renormalization Group (DMRG) \cite{chan2011density} calculation and use Matrix Product State (MPS) circuit synthesis \cite{schon2005sequential, berry2025rapid}. While these methods are robust and versatile, they require rather deep quantum circuits, due to the heavy use of multi-controlled rotation gates. On the other hand, variational methods based on the Variational Quantum Eigensolver (VQE) \cite{peruzzo2014variational} can be used to construct shallower circuits: even the Efficient Symmetry Ansatz (which can parameterize any quantum state) \cite{gard2020efficient} results in about an order of magnitude lower number of quantum gates than CIPSI with CVO-QRAM, for hydrogen chains up to \ce{H14}~\cite{feniou2024sparse}. An adaptive method for Unitary Coupled Cluster ansatz construction can deliver even larger savings: Overlap-ADAPT-VQE yielded the same state fidelity for \ce{H14} as CIPSI/CVO-QRAM at $10^3$ lower gate cost \cite{feniou2024sparse}. It must be stressed, however, that VQE-based methods represent a challenging optimization problem \cite{larocca2024review, astrakhantsev2023phenomenological}, and suffer from prohibitive scaling of sampling cost with system size when executed on actual quantum computers \cite{gonthier2022measurements}. 
In this work, we introduce oo-pCCD-UpCCD, a low-cost state preparation method for singlet electronic states, yielding high-fidelity states without the need for variational optimization or deep circuits. We demonstrate the utility of our approach for determination of ground state energy of stretched ethene molecule using Quantum Phase Estimation.

\section{Theory}
\subsection{Seniority-zero methods for singlet states}
Let us consider the special case of strongly-correlated (or `multi-configurational') singlet states. It is well-known that Doubly Occupied Configuration Interaction (DOCI, seniority-zero CI), defined as a linear combination of Slater determinants in which every spatial orbital is either unoccupied or doubly occupied, yields (when used with self-consistently optimized orbitals) an excellent  description of static correlation in strongly-correlated systems with a spin of zero \cite{limacher2013new, bytautas2011seniority,kossoski2022hierarchy, couty1997generalized}. Orbital-optimized paired Coupled Cluster Doubles (oo-pCCD) theory provides an inexpensive ($\mathcal{O}(N^2)$ amplitudes, $\mathcal{O}(N^3)$ time complexity of pCCD) yet surprisingly accurate approximation to DOCI especially for molecular Hamiltonians \cite{henderson2014seniority}, e.g. for dissociation of \ce{H4} - \ce{H8} \cite{limacher2013new, henderson2014seniority}, dissociation of \ce{LiH}, \ce{H2O} and \ce{N2}\cite{henderson2014seniority} as well as the 1-D \cite{stein2014seniority} and 2-D Hubbard models \cite{shepherd2016using}. Thus, pCCD appears to be an attractive starting point for designing a compact representation for singlet initial states. We stress that the orbital optimization is essential for the success of pCCD in approximating DOCI\cite{stein2014seniority}.

\subsection{Coupled Cluster methods in Quantum Circuits}
Standard Coupled Cluster theories do not lend themselves to a direct representation in the form of a quantum circuit, because the exponentiated cluster operator $e^{\hat{T}}$ is non-unitary, where $\hat{T}$ is a general excitations operator. On the other hand, Unitary Coupled Cluster (UCC), which uses a unitary $e^{\hat{T}-\hat{T}^\dagger}$ operator instead, when truncated (e.g. at single and double excitations, as UCCSD) and approximated with a product formula forms a popular family of parameterized circuits used in VQE \cite{anand2022quantum}. Quantum circuits implementing Unitary pCCD (UpCCD) have been introduced by Elfving \latin{et al.} in the context of a `bosonic' mapping \cite{elfving2021, elfving2021simulating, o2023purification} and generalized to Jordan-Wigner mapping by Nam \latin{et al.}\cite{nam2020ground}. Thanks to the fact that paired double excitations obey bosonic statistics, there is no need to track state parity when transforming to the qubit basis, and each $e^{\hat{T_{a}^{i}}-\hat{T_{a}^{i}}^\dagger}$ term is implemented by a Givens rotation, followed by a layer of CX gates \cite{nam2020ground}, in stark contrast to up to 8 \textit{Pauli gadgets} (each consisting of a double CX ladder sandwiching a parameterized rotation) required for standard UCCD \cite{cowtan2020generic}. 

\subsection{Exact relations between CC, disentangled UCC and CI}
The relation between trotterized (`disentangled') UCCSD, traditional CCSD and CI has been derived by Evangelista \cite{evangelista2019exact} for the simple case of two electrons in four spinorbitals:
$\ket{\psi} = c_1 \ket{1100} 
    + c_2 \ket{0110} 
    + c_3 \ket{1001} 
    + c_4 \ket{0011}.$ While the relation to CCSD is simple: 
\begin{align}
\Psi_{\rm{CCSD}} &= e^{t_1\hat{a}_2^\dagger\hat{a}_0
+t_1'\hat{a}_3^\dagger\hat{a}_1
+t_2\hat{a}_2^\dagger\hat{a}_3^\dagger\hat{a}_1\hat{a}_0}\ket{1100}\\
c_1 &= 1, c_2 = t_1, c_3 = t_1', c_4 = t_2+ t_1t_1', 
\end{align}
the CI coefficients corresponding to one choice of operator ordering in disentangled UCCSD are \cite{evangelista2019exact}:
\begin{align}
   \Psi_{\rm{UCCSD}}&= e^{t_1'(\hat{a}_3^\dagger\hat{a}_1-\hat{a}_1^\dagger\hat{a}_3)}
e^{t_2(\hat{a}_2^\dagger\hat{a}_3^\dagger\hat{a}_1\hat{a}_0
   -\hat{a}_0^\dagger\hat{a}_1^\dagger\hat{a}_3\hat{a}_2)}
   e^{t_1(\hat{a}_2^\dagger\hat{a}_0-\hat{a}_0^\dagger\hat{a}_2)}\ket{1100}\\
    c_1 &= \cos(t_1) \cos(t_1') \cos(t_2) \nonumber \\
    c_2 &= \sin(t_1) \cos(t_1') \cos(t_1) \sin(t_1') \sin(t_2) \nonumber \\
    c_3 &= \cos(t_1) \sin(t_1') \cos(t_2) \nonumber \\
    c_4 &= \sin(t_1) \sin(t_1') + \cos(t_1) \cos(t_1') \sin(t_2),
\end{align}
demonstrating that, in general, CCSD and UCCSD amplitudes for the same state should be quite different. A general recipe for deriving the action of each term in disentangled UCC on a quantum state has been provided by Chen \latin{et al.} \cite{chen2021quantum}, and is shown below for single and double excitation terms: 
\begin{align}
\exp[\theta (\hat{a}_a^\dagger\hat{a}_i-\hat{a}_i^\dagger\hat{a}_a)] = 1 &+ \sin\theta(\hat{a}_a^\dagger\hat{a}_i-\hat{a}_i^\dagger\hat{a}_a) \nonumber \\
 &+ (\cos\theta -1)(\hat{n}_a + \hat{n}_i - 2\hat{n}_a\hat{n}_i),  \\
\exp[\theta(\hat{a}_{a_1}^\dagger \hat{a}_{i_1} \hat{a}_{a_2}^\dagger \hat{a}_{i_2}- \hat{a}_{i_2}^\dagger\hat{a}_{a_2} \hat{a}_{i_1}^\dagger\hat{a}_{a_1})] = 1 
&+ \sin\theta (\hat{a}_{a_1}^\dagger \hat{a}_{i_1} \hat{a}_{a_2}^\dagger \hat{a}_{i_2}- \hat{a}_{i_2}^\dagger\hat{a}_{a_2} \hat{a}_{i_1}^\dagger\hat{a}_{a_1}) \nonumber \\
&+(\cos\theta-1)[\hat{n}_{a_1}\hat{n}_{a_2}(1-\hat{n}_{i_1})(1-\hat{n}_{i_2}) \nonumber \\
&+(1-\hat{n}_{a_1})(1-\hat{n}_{a_2})\hat{n}_{i_1}\hat{n}_{i_2}], \label{eq:uccd}
\end{align}
where $\theta$ is a cluster amplitude, $i$ and $a$ are, respectively, indices of orbitals occupied and unoccupied in the reference state, and $\hat{n}_\alpha=\hat{a}_\alpha^\dagger\hat{a}_\alpha$ is the number operator. 

\subsection{Relations between pCCD and UpCCD}
Motivated by the effectiveness and low cost oo-pCCD calculations, and the desire to avoid hybrid quantum-classical variational optimization or deep QRAM-type circuits, we set out to explore connections between non-unitary pCCD and trotterized UpCCD (henceafter referred to as UpCCD). Applying the small angle approximation [$\sin \theta = \theta + \mathcal{O}(\theta^3), \cos \theta = 1 + \mathcal{O}(\theta^2)$, i.e. Taylor series expansion truncated above linear terms] to Eq.~(\ref{eq:uccd}) and taking $i_2=i_1+1, a_2 = a_1+1$ (paired excitations), we obtain:
\begin{align}
\exp[\theta(\hat{a}_{a}^\dagger \hat{a}_{i} \hat{a}_{a+1}^\dagger \hat{a}_{i+1}- \hat{a}_{i+1}^\dagger\hat{a}_{a+1} \hat{a}_{i}^\dagger\hat{a}_{a})] \approx 1 
&+ \theta (\hat{a}_{a}^\dagger \hat{a}_{i} \hat{a}_{a+1}^\dagger \hat{a}_{i+1}- \hat{a}_{i+1}^\dagger\hat{a}_{a+1} \hat{a}_{i}^\dagger\hat{a}_{a}), \label{eq:sa-upccd}
\end{align}
where we dropped the subindices for simplicity of notation. Thus, the action of each term in UpCCD can be approximated (in the limit of small amplitudes) as the sum of identity, a double excitation and a double deexcitation. The error of the small-angle approximation is, due to the dominance of the error of $\cos \theta = 1+ \frac{\theta^2}{2} + \dots\approx 1$, quadratic in the amplitudes $\theta$ and can be bounded by $n_{amps}\frac{\theta_{max}^2}{2}$, where $n_{amps}$ is the number of non-zero cluster amplitudes and $\theta_{max}$ is the largest amplitude.

To illustrate the similarities between pCCD and UpCCD, we first write the pCCD wave function for 4 electrons in 4 orbitals (8 spinorbitals) ($t_{o,v}$ is the amplitude of the excitation of an electron pair from  orbital $o$ to $v$):
\begin{align}
\Psi_{(4,4)-\rm{pCCD}} &= \ket{11110000} + t_{1,2} \ket{11001100} 
+ t_{1,3}\ket{11000011} \nonumber \\ 
&+ t_{0,2}\ket{00111100} + t_{0,3}\ket{00110011} \nonumber \\
&+ (t_{1,2}t_{0,3} + t_{1,3}t_{0,2})\ket{00001111}. \label{eq:pccd_4_8}
\end{align}
The corresponding UpCCD wave function, assuming an operator ordering resulting from iterating first over $o$, then over $v$, can be approximated by repeated application of Eq. ~(\ref{eq:sa-upccd}) as:
\begin{align}
\Psi_{(4,4)-\rm{UpCCD}} &\approx \ket{11110000} + t_{1,2} \ket{11001100} 
+ t_{1,3}\ket{11000011} \nonumber \\ 
&+ (t_{0,2}-t_{1,2}t_{0,3}t_{1,3})\ket{00111100} + t_{0,3}\ket{00110011} \nonumber \\
&+ (t_{1,2}t_{0,3} + t_{1,3}t_{0,2})\ket{00001111}. \label{eq:upccd_4_8}
\end{align}
We notice that Eq.~(\ref{eq:pccd_4_8}) and (\ref{eq:upccd_4_8}) differ only in the coefficient of the $\ket{00111100}$ determinant, by $-t_{1,2}t_{0,3}t_{1,3}$: a term generated by two double excitations and one double deexcitation, which we call a 3-chain of interacting excitations. Being a product of three amplitudes (each being smaller than 1), this deviation from the corresponding CI coefficient in the pCCD ($=t_{0,2}$) is likely to be much smaller than the $t_{0,2}$ itself and must vanish if any of the numbers $t_{1,2}$, $t_{0,3}$ or $t_{1,3}$ are zero, rendering the small-angle-approximated UpCCD and pCCD wave functions equal, given the same amplitudes in both Ans\"atze. We note that the presence of such terms depends on the ordering of operators in the product formula, and e.g. reversing the operators in the above UpCCD wave function leads to this term appearing in the coefficient for $\ket{11000011}$, and it may be possible to minimize the effect of those product terms by choosing a special ordering of the operators (which we do not investigate herein). Furthermore, UpCCD wave functions of systems with more electrons may contain higher terms generated by chains (products) of $n$ double excitations and $n-1$ double deexcitations (up to $2 \leq n \leq \frac{e_{max}}{2}$, where $e_{max}$ is the highest excitation possible in the system).

\begin{figure}[ht]
\includegraphics{"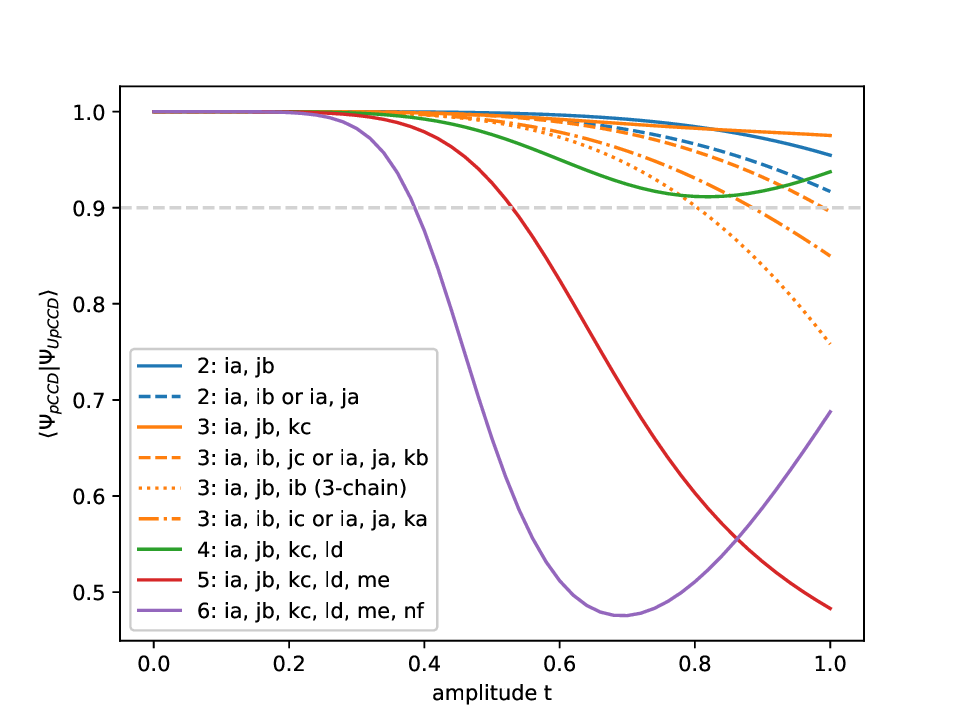"}
\caption{Overlaps of UpCCD and pCCD wave functions sharing the same amplitudes, for different numbers of non-zero amplitudes and various interactions between excitations (see text)} \label{fig:pccd_upccd_overlaps}
\end{figure}

However, if the magnitudes of the amplitudes are larger than ca. $0.2-0.3$, the small angle approximation no longer holds. For CC wave functions with a small number of non-zero amplitudes, we can derive analytical formulae for $\braket{\Psi_{\rm{pCCD}}}{\Psi_{\rm{UpCCD}}}$ as a function of the amplitudes and the interactions between excitations. We find that  these overlaps depend neither on the number of orbitals or electrons in the system. For example, the overlap of pCCD and trotterized UpCCD wave functions having two non-zero amplitudes $t_{ia}$ and $t_{jb}$ (and, for simplicity, $t=t_{ia}=t_{jb}$) which do not interact (i.e. $i\neq j, a \neq b$) equals:
\begin{align}
\braket{\Psi_{\rm{pCCD}}}{\Psi_{\rm{UpCCD}}}_{ia,jb}=\frac{
t^{2} \sin^{2}{t} + t \sin{2t} + \cos^{2}{t}}
{t^{2} + 1}, \label{eq:pccd_upccd_overlap_iajb}
\end{align}
whereas if those two amplitudes are interacting via an occupied or virtual orbital ($t=t_{ia}=t_{ib}$ or $t=t_{ia}=t_{ja}$), the overlap becomes:
\begin{align}
\braket{\Psi_{\rm{pCCD}}}{\Psi_{\rm{UpCCD}}}_{ia,ib}=\braket{\Psi_{\rm{pCCD}}}{\Psi_{\rm{UpCCD}}}_{ia,ja}=\frac{t \sin{t} + \frac{t \sin{2t}}{2} + \cos^{2}{t}}{\sqrt{2 t^{2} + 1}}.
\end{align}
We explain the derivation and show analogous formulae for overlaps of wave functions with 3-6 non-interacting amplitudes, as well as those with an interacting chain of 3 amplitudes, in the Supporting Information. We find that $\braket{\Psi_{\rm{pCCD}}}{\Psi_{\rm{UpCCD}}}$ is large (i.e. above 0.9) if the magnitudes of amplitudes are below ca. 0.3 (i.e. the small angle approximation is valid) or if the number of non-zero amplitudes does not exceed 4, see Fig.~\ref{fig:pccd_upccd_overlaps}. The effect of 3-chains of amplitudes, which give rise to errors at the small-angle approximation level, is substantial (dotted orange line), but it is the sheer presence of more non-zero amplitudes which breaks the approximation. 
We note that using CCSD amplitudes as an initial guess to UpCCD parameters (prior to optimization with VQE) was previously reported by Elfving \cite{elfving2021simulating}, but no theoretical justification was given. However, it is well-known that CCSD often does not correctly describe dissociation curves, therefore their approach is unlikely to provide a good initial estimate of UpCCD angles when CCSD fails, for example in the case of dissociation of \ce{N2} (see Supporting Information).

In the following sections, we show numerical experiments in which substituting largest oo-pCCD amplitudes into an UpCCD ansatz allows preparation of states with high overlaps with the ground states with very shallow circuits, for a range of systems for which the Hartree-Fock wave function has low overlap with the ground state. Furthermore, we demonstrate the utility of our method in the Quantum Phase Estimation algorithm. 

\section{Calculations}
Mean-field calculations, preparation of 1- and 2-electron integrals and CASCI calculations have been performed with PySCF \cite{sun2020recent}. Orbital-optimized pCCD has been carried out using an in-house code (partially based on PySCF), using automatic differentiation as implemented in Jax \cite{jax2025github} and the Adam \cite{kingma2017adammethodstochasticoptimization} optimizer from Optax \cite{deepmind2020jax}, modified to control the step size via an adaptive trust radius, for the orbital optimization. Quantum circuits were constructed using an in-house code, built on top of pytket \cite{sivarajah2020t}. UpCCD circuits were constructed using Givens rotation gates as described by Nam \latin{et al.} \cite{nam2020ground}, but with an opposite sign convention for the parameters, for consistency with pCCD (see Supporting Information). Canonical Quantum Phase Estimation simulations used a first-order product formula approximation for the evolution operator, with a scaled Hamiltonian (see Supporting Information) and the eigenphase estimate was obtained from the most abundant measurement outcome and converted back to the energy. We used Qulacs \cite{suzuki2021qulacs} for state vector circuit simulation, and pytket-cutensornet \cite{pytket_cutensornet} for estimation of state overlap in cases involving more than 20 qubits. Comparative experiments with CVO-QRAM \cite{deveras2022double, qclib} were performed with our own implementation, using multi-controlled gate decompositions as implemented in pytket 2.0 \cite{sivarajah2020t} (see SI for details). Reported circuit depths refer to circuits  optimized using the FullPeepholeOptimise pass in pytket \cite{sivarajah2020t} and rebased to a common gate set (see Supporting Information). For details of UpCCD circuit construction, as well as all UpCCD and CVO-QRAM circuits, see Supporting Information. Wherever we refer to an equivalent-fidelity CVO-QRAM circuit, we mean a circuit which prepares a state defined by as many CI determinants taken from the seniority-zero sector of the CASCI or HCI expansion, starting from those with largest coefficient magnitudes, as required to reach at least the same fidelity as that of the UpCCD state.  Heat-bath Configuration Interaction (HCI) calculations were performed using Dice \cite{sharma2017semistochastic,holmes2016heat,smith2017cheap}, with $\epsilon$ of $1\times 10^{-3}$ -- $8 \times 10^{-4}$, and 20000 largest CI coefficients were used for estimation of overlap with the UpCCD states.

\section{Numerical results and discussion}
\subsection{1-D Hubbard Model}
\begin{figure}[ht]
\includegraphics{"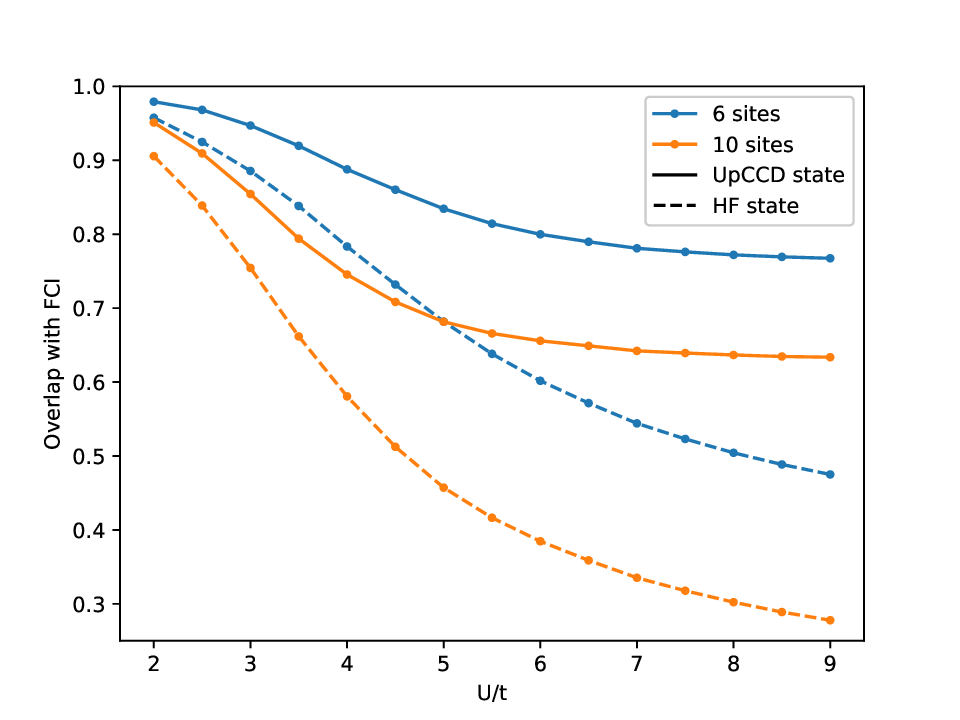"}
\caption{Overlaps of UpCCD (solid lines) and Hartree-Fock (dashed lines) states with CASCI states, for 6-, 10- and 12-site Hubbard models at half filling} \label{fig:hubbard_overlaps}
\end{figure}

The Hubbard Hamiltonian, a widely-studied model system, is a particularly useful test bed for electronic structure methods, because its properties can be continuously tuned from non-interacting to strongly correlated\cite{scalettar2016introduction}. While oo-pCCD does not exactly reproduce the eigenstates of 1D Hubbard chains at half-filling, it captures the static correlation effects very well, even at high values of U/t \cite{stein2014seniority}. Therefore, we investigated the overlap of exact eigenstates with UpCCD wave functions (with amplitudes from oo-pCCD), for half-filled 1D Hubbard models with 6 and 10 sites (corresponding to 12 and 20 qubits, respectively), varying U from 0.5t to 9t, see Fig. \ref{fig:hubbard_overlaps}. To limit the circuit depth, only amplitudes having magnitudes larger than 0.1 have been used in the UpCCD Ansatz (see Supporting Information for a detailed analysis).  Across the tested range of U, the UpCCD wave functions have consistently higher overlaps with the ground state than the HF wave functions. For low values of U, the overlaps are very high, while at high U the overlap plateaus at a value dependent on the chain length. However, overlaps of UpCCD with the seniority-zero sector of FCI remain above 0.99 even for high values of U (see Fig. S14), demonstrating that the loss of fidelity is due to neglect of higher seniority determinants. Our state preparation circuits are very shallow: 8-10 gates in depth. In contrast, for $U/t=8$, equivalent-fidelity CVO-QRAM circuits have depths of up to 892 and 6232 gates, respectively (see Fig. S9 and S13).

\subsection{Bond dissociation in ethene, ethyne and dinitrogen}

Dissociation of ethene \ce{C2H4} into two triplet carbenes is one of the simplest reactions which require a multi-configurational wave function even for a qualitatively correct description \cite{ohta1985dimerization}. Dissociation of dinitrogen \ce{N2} \cite{Chan2004} and ethyne \ce{C2H2} \cite{Siegbahn1981,zeng2014tuning} are also well-studied multi-configurational problems. For each of these systems, we varied the length of the multiple bond, while the remaining internal coordinates, except for \ce{N2}, were taken from the W4-11 data set \cite{karton2011w4} (see the Supporting Information for xyz coordinates); and performed oo-pCCD in the cc-pVDZ basis \cite{Dunning1989}. 
\subsubsection{Bond dissociation: 20-qubit models}
\begin{figure}[ht]
\includegraphics{"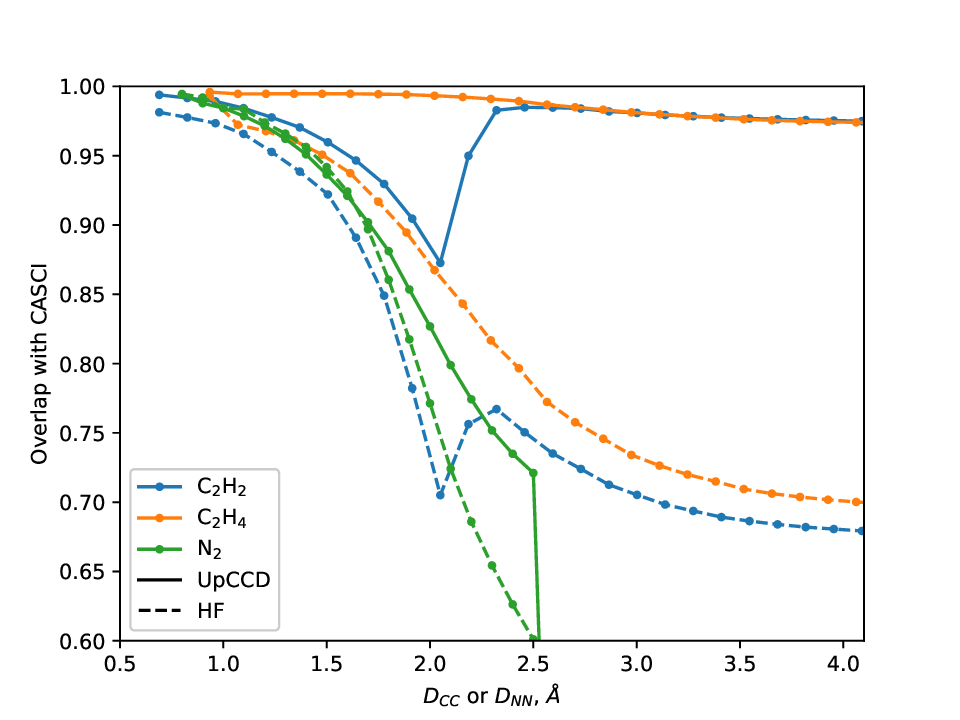"}
\caption{Overlaps of UpCCD (solid lines) and Hartree-Fock (dashed lines) states with CASCI states, for 20-qubit models of \ce{C2H2}, \ce{C2H4} and \ce{N2}, vs. C-C or N-N distances (see text)} \label{fig:upccd_20q_overlaps}
\end{figure}

For an analysis of the fidelity of our oo-pCCD-UpCCD scheme along the bond dissociation curves, we have chosen an active space of 10 electrons in 10 (spatial) orbitals: large enough to capture static correlation effects, but small enough to enable rapid calculations. For each geometry, oo-pCCD was calculated in the 10-electron, 10-orbital active space and a 20-qubit UpCCD ansatz was constructed using oo-pCCD amplitudes having magnitudes larger than 0.01 (\ce{C2H2} and \ce{C2H4}) or 0.1 (\ce{N2}), see Supporting Information for threshold sensitivity analysis. The Ansatz was simulated using Qulacs~\cite{suzuki2021qulacs} and the overlap with the lowest-lying singlet CASCI state (obtained for the same active space and orbitals) was computed. Fig. \ref{fig:upccd_20q_overlaps} compares the fidelities of such-constructed UpCCD states to those of Hartree-Fock (HF) states (note that it is possible to obtain higher fid. As expected, around equilibrium geometries ($D_{CC}= 1.206 $ \text{\AA} for \ce{C2H2}, $D_{CC}= 1.232$ \text{\AA} for \ce{C2H4} and $D_{NN}=1.190$ \text{\AA} for \ce{N2}), both UpCCD and HF states have near-perfect overlaps with the exact ground states. However, when the multiple bonds are stretched, the quality of the HF solution deteriorates more rapidly than that of the UpCCD with pCCD amplitudes, although the behaviour is markedly different for each system. In the case of \ce{C2H2}, UpCCD has modest advantage over HF up to ca. 2.2-2.3 \text{\AA} (perhaps due to the avoided crossing, see Fig. 4 in Zeng \latin{et al.}\cite{zeng2014tuning}), above which the accuracy of the UpCCD ansatz improves to near-perfect. For ethene, we observe a smooth, very slight decrease of UpCCD's fidelity during dissociation, from near-1 to about 0.97, while overlap with the HF state decreases to about 0.7 at 2.84 \text{\AA} and drops to zero at longer distances. In the case of \ce{N2}, the gap between HF's and UpCCD's fidelities grows steadily with bond length, despite a decrease in UpCCD's quality. At $D_{NN}=2.5 $ \text{\AA}, UpCCD has overlap of 0.7 with FCI while at larger internuclear separations our Ansatz is no longer adequate. Interestingly, the decrease of UpCCD state fidelity with increasing bond length for \ce{N2} and in the intermediate bond length regime for \ce{C2H2} can be attributed mainly to neglect of higher seniority determinants (i.e. non-paired excitations) rather than discrepancies between UpCCD and pCCD, because overlaps between UpCCD and seniority-zero sector of CASCI are consistently high (see Fig. S24).

\subsubsection{State fidelity in all-electron calculations}
We applied the oo-pCCD-UpCCD scheme, with amplitude threshold of 0.1 (i.e. leading 3 amplitudes), to all-electron calculations of \ce{C2H2} at $D_{CC} = 2.38$ \text{\AA} (i.e. $2.0 \times D_{eq}$) and \ce{N2} at $D_{NN}=2.5$ \text{\AA} (i.e. $2.1 \times D_{eq}$), in cc-pVDZ and cc-pVTZ basis sets. Due to the number of qubits exceeding the capabilities of a state vector simulator, simulations were performed with the pytket-cutensornet tensor network simulator, and the exact ground states were approximated with Heat-Bath Configuration Interaction. The results are shown in Table \ref{tbl:all-electron}, additional details are presented in the Supporting Information. In each case, states prepared with very shallow circuits (depth of 10 gates) have very high (stretched ethyne) or, at least, decent, i.e. $2\times$ better than HF (stretched dinitrogen), overlaps with HCI. For comparison, we have also constructed CVO-QRAM circuits, preparing states defined by as many of the CI coefficients with the largest absolute values as required to achieve a fidelity no smaller than that of the corresponding UpCCD state. The differences are striking: oo-pCCD-UpCCD required two orders of magnitude lower depth than CVO-QRAM. 
\begin{table}
  \caption{Overlaps of oo-pCCD-UpCCD states with exact ground states of stretched \ce{C2H2} and \ce{N2} molecules}
  \label{tbl:all-electron}
  \begin{tabular}{ccccccc}
    \hline
    System  &  Basis & $N_q$ & $ \braket{\Phi_{\rm{UpCCD}}}{\Phi_{\rm{HCI}}} $ & Depth & $ \braket{\Phi_{\rm{HF}}}{\Phi_{\rm{HCI}}} $  & CVO-QRAM depth  \\
    \hline
    \ce{C2H2}  & cc-pVDZ & 76  & 0.95 & 10 & 0.76 &  1377  \\
    \ce{C2H2}  & cc-pVTZ & 176 & 0.95 & 10 & 0.76 &    \\
    \ce{N2}    & cc-pVDZ & 56  & 0.69 & 10 & 0.34 &   2182  \\
    \ce{N2}    & cc-pVTZ & 120 & 0.69 & 10 & 0.34 &     \\
    \hline
  \end{tabular}

\end{table}

\subsubsection{Application to Quantum Phase Estimation}
\begin{figure}[ht]
\includegraphics{"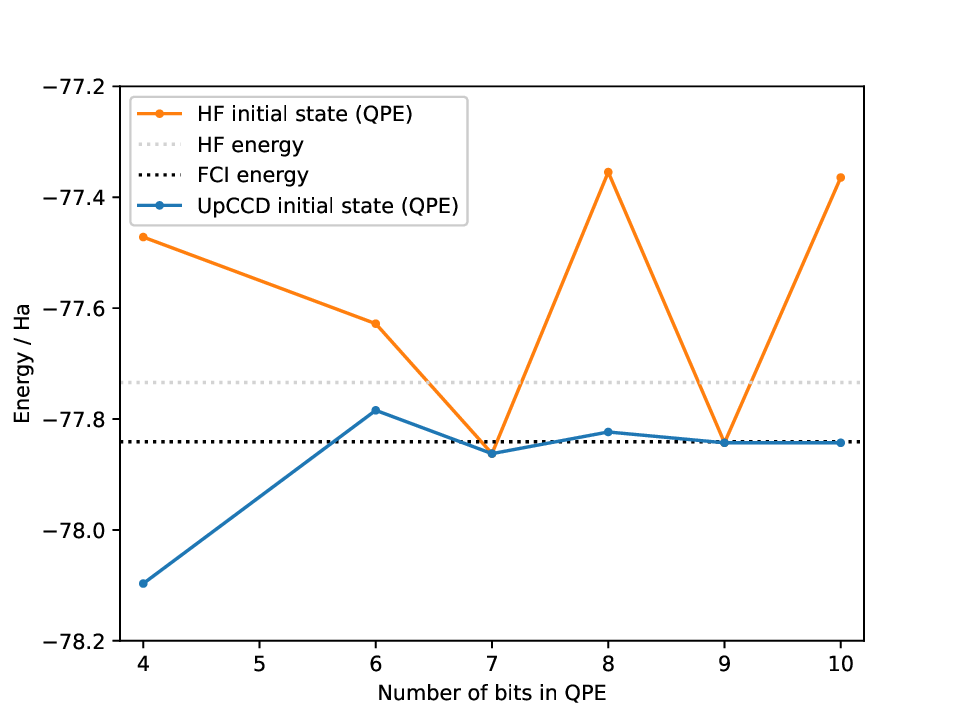"}
\caption{Effect of the initial state on the outcome of Canonical   Quantum Phase Estimation applied to singlet state of dissociated \ce{C2H4}} \label{fig:c2h4_qpe_conv}
\end{figure}

\begin{figure}[h]
\includegraphics{"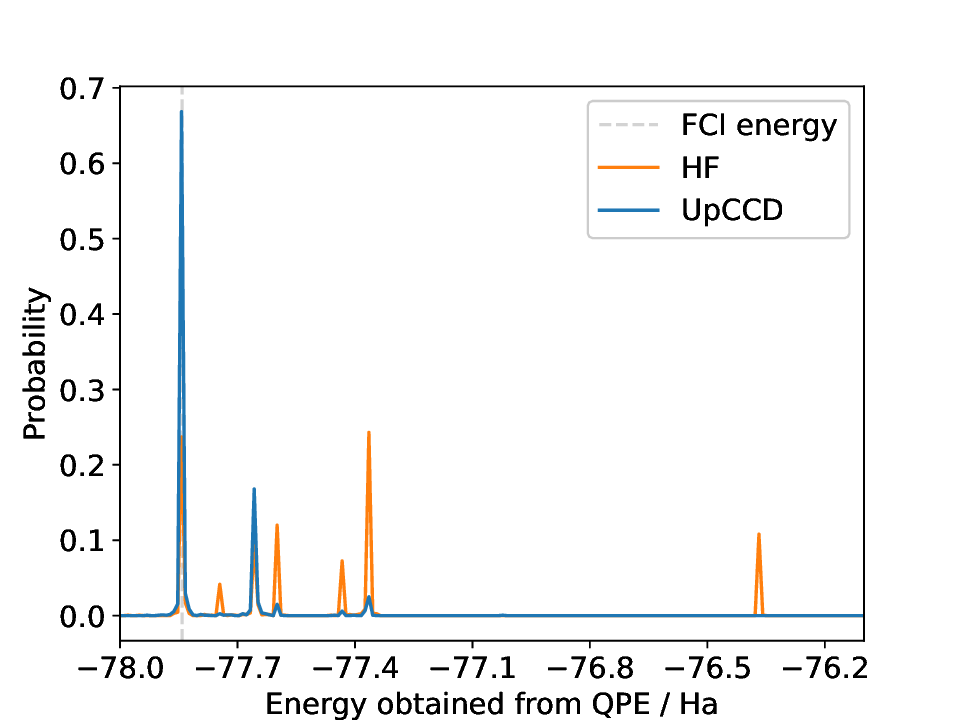"}
\caption{Effect of the initial state on the outcome of Canonical Quantum Phase Estimation applied to singlet state of dissociated \ce{C2H4}: histogram of energies (2000 samples)} \label{fig:c2h4_qpe_hist}
\end{figure}

To demonstrate the usefulness of the oo-pCCD-UpCCD scheme for state preparation for QPE, we have revisited the ethene dissociation reaction mentioned above. We constructed a minimal model of the double bond dissociation in \ce{C2H4}, using the geometry as reported in the W4-11 data set \cite{karton2011w4} with the C-C distance set to 3.23 \text{\AA} (i.e. $D_{eq}+2$ \text{\AA}). Note that the lowest singlet state at this geometry corresponds to two coupled triplet carbenes \cite{ohta1985dimerization}. We performed oo-pCCD, restricting the orbital rotation to the active space and starting from (4,4)-CASSCF orbitals. We construct the UpCCD Ansatz by plugging the two largest pCCD amplitudes ($t_{0,3} = -0.6940$ and $t_{1,2} = -0.9270$) into the Ansatz. Success probability of QPE for this system with a single-determinantal (HF) state in the oo-pCCD orbital basis is expected to be $\braket{\Psi_{\rm{HF},\rm{oo-pCCD}}}{\Psi_{\rm{CASSCF}}}^2=0.28$ while UpCCD gives a much higher probability $\braket{\Psi_{\rm{UpCCD}}}{\Psi_{\rm{CASSCF}}}^2=0.75$. To demonstrate this numerically, we performed Canonical Quantum Phase Estimation (with up to 10 ancilla registers) on the shifted and rescaled Hamiltonian in oo-pCCD orbitals, using the UpCCD or HF states as the initial states, see Fig. \ref{fig:c2h4_qpe_conv}. Clearly, the HF state is insufficiently close to the ground state to ensure success, while QPE applied to the UpCCD initial state reliably converges to the ground state energy. The impact of the state preparation method is even more evident when we plot the histogram of energies obtained from 10-ancilla canonical QPE sampled 2000 times (Fig. \ref{fig:c2h4_qpe_hist}): initialization with UpCCD yields a large peak (0.73 of the total counts) corresponding to the ground state and 2-3 smaller peaks of excited states. Starting from the HF state results in a complicated spectrum, in which the highest peak is 0.48 Ha above the ground state while the ground state peak is relatively small (0.26 of the total counts). 

\subsection{Conclusions and Outlook}
In this work, we have derived the connection between oo-pCCD and UpCCD wave functions showing that overlaps between these two Ansatze are high if amplitudes have small magnitudes or if the number of non-zero amplitudes is small. We have introduced the oo-pCCD-UpCCD ansatz as a non-variational approximate state preparation method for singlet states, for use with Quantum Phase Estimation. We evaluated the effectiveness of our scheme in preparing high-fidelity states, on prototypical multi-configurational examples of double bond dissociation (ethene) and triple bond dissociation (ethyne and dinitrogen), as well as on a model Hamiltonian with correlation tunable from weak to strong (1D Hubbard models with 6 and 10 sites). In each case, the UpCCD ansatz circuits with leading oo-pCCD amplitudes were embarrassingly shallow and prepared states with significantly higher fidelities than the Hartree-Fock state (apart from \ce{N2} with $D_{NN}<1.6$, where overlaps of UpCCD and HF with HCI were similar and $>0.9$); high enough to ensure success of Quantum Phase Estimation. We show, on the stretched \ce{C2H2} and \ce{N2} examples, that the pCCD state has almost unit overlap with the seniority-zero sector of HCI, and that the overlap between pCCD and UpCCD states is high, owing to a small number of large amplitudes. Finally, we demonstrate how our scheme enables successful estimation of ground state energy of an 8-qubit model of dissociated ethene.

Overall, this study enhances our understanding of the connections between non-unitary and unitary coupled cluster methods with paired doubles, and provides a low cost method for preparing high-fidelity singlet states, for use with QPE and related methods.







\bibliography{references}

\end{document}